\documentclass[letter]{aa}

\usepackage{graphicx}
\usepackage{txfonts}

\begin{document}

\title{A remarkable recurrent nova in M31 - The optical observations}

\author{M. J. Darnley\inst{1} \and S. C. Williams\inst{1} \and
  M. F. Bode\inst{1} \and M. Henze\inst{2} \and J.-U. Ness\inst{2}
  \and A. W. Shafter\inst{3} \and K. Hornoch\inst{4} \and V. Votruba\inst{4}}

\institute{Astrophysics Research Institute, Liverpool John Moores
  University, IC2 Liverpool Science Park, Liverpool, L3 5RF, UK \\
\email{M.J.Darnley@ljmu.ac.uk} \and European Space Astronomy Centre, P.O.\ Box 78, 28691 Villanueva
  de la Ca\~{n}ada, Madrid, Spain \and Department of Astronomy, San Diego State University, San Diego,
  CA 92182, USA \and Astronomical Institute, Academy of Sciences,
  CZ-251 65 Ond\v{r}ejov, Czech Republic}

\date{Received ?; accepted ?}

\abstract{In late November 2013 a fifth eruption in five years of the M31 recurrent
nova M31N~2008-12a was announced.}
{In this Letter we address the optical lightcurve and progenitor
  system of M31N~2008-12a.}
{Optical imaging data of the 2013 eruption from the Liverpool
  Telescope, La Palma, and Danish 1.54m Telescope, La Silla, and archival {\it Hubble Space
    Telescope} near-IR, optical and near-UV data are astrometrically
  and photometrically analysed.}
{Photometry of the 2013 eruption, combined with three previous eruptions,
  enabled construction of a template lightcurve of a very fast
  nova, $t_{2}\left(V\right)\simeq4$ days.  The archival data allowed recovery of the
  progenitor system in optical and near-UV data, indicating a
  red-giant secondary with bright accretion disk, or alternatively a
  system with a sub-giant secondary but dominated by a disk.}
{The eruptions of M31N~2008-12a, and a number of historic
  X-ray detections, indicate a unique system with a recurrence
  timescale of $\sim1$ year.  This implies the presence of a very
  high mass white dwarf and a high accretion rate.  The recovered
  progenitor system is consistent with such an elevated rate of
  accretion.  We encourage additional observations, especially towards
  the end of 2014.}


\maketitle

\section{Introduction}

Classical and Recurrent Novae (CNe \& RNe) are cataclysmic variable
stars that exhibit eruptions driven by a thermonuclear runaway on
the surface of a white dwarf (WD; the primary) in an interacting binary system.
CN systems typically contain main sequence secondaries with
expected recurrence times of a few~$\times10^{3}-10^{6}$~years
\citep[see][for recent reviews]{2008clno.book.....B}.  RNe
are observed to recur on timescales of ten to a hundred years and most
contain evolved, sub-giant or red giant, secondaries \citep[SG-
and RG-novae respectively;][]{2012ApJ...746...61D}. 

While hundreds of Galactic CNe are known, there are only ten
confirmed RNe in the Milky Way.  With
over 900 novae discovered in \object{M31} \citep[and on-line
database\protect{\footnote{http://www.mpe.mpg.de/\textasciitilde
    m31novae/opt/m31/index.php}}]{2007A&A...465..375P} and with a nova rate
of $65^{+16}_{-15}\;\mathrm{yr}^{-1}$\citep{2006MNRAS.369..257D} \object{M31} remains a potentially huge untapped resource for identifying
RNe with only a handful of candidate systems known.

The RN \object{M31N 2008-12a} has been discovered in eruption
optically five times over a five year period, in 2008, 2009, 2011, 2012 and
most recently in Nov 2013.   For comparison, the shortest
inter-eruption time for a Galactic RN is 8 years in the case of
\object{U Sco} \citep{2010ApJS..187..275S}.  Eruptions of the system have
also been detected in X-rays at entirely separate epochs, an
overview of which is given in the accompanying Letter by
\citep[hereafter HND2014]{2014arXiv1401.2904H}.  Here we briefly describe
the optical discoveries up to and including the 2012 eruption, the
Dec 2009 eruption is described in full in \citet{2014arXiv1401.2426T}. 

K. Nishiyama and F. Kabashima discovered a \object{M31} nova candidate
(\object{M31N 2008-12a}) with an
unfiltered magnitude of 18.7 at
$0^{\mathrm{h}}45^{\mathrm{m}}28^{\mathrm{s}}\!.80$,
$+41^{\circ}54^{\prime}10^{\prime\prime}\!\!.1$ (J2000) in images
taken on 2008 Dec 26.48 UT\footnote{http://www.cbat.eps.harvard.edu/CBAT\_M31.html\#2008-12a}.  Liverpool
Telescope \citep[LT;][]{2004SPIE.5489..679S} RATCam observations taken
on 2009 Jan 10.85 showed no resolvable objects at the nova
position down to a $B$-band limiting magnitude of 19.9 and
$B>21.2$ on 2009 Jan 13.93.  The nova was not visible at any other
wavebands down to the following limiting magnitudes: 20.1 on Jan 10.85
and 21.2 on Jan 13.93 in $V$-band, 20.8 on Jan 10.84 and 21.0 on
Jan 13.92 in $r'$-band, 19.8 on Jan 10.84 and 20.9 on Jan 13.92 in
$i'$-band.

In 2011 an eruption (\object{M31N 2011-10e}) coincident with the position of
\object{M31N 2008-12a} was discovered by
S. Korotkiy and L. Elenin in data taken on 2011 Oct
22.46 UT at an unfiltered magnitude of $18.6\pm0.3$, they measured the position as
$0^{\mathrm{h}}45^{\mathrm{m}}28^{\mathrm{s}}\!.85$,
$+41^{\circ}54^{\prime}09^{\prime\prime}\!\!.4$
($\pm0^{\prime\prime}\!\!.3$).  They reported that no object
was visible on Oct 21.35 to limiting magnitude of
$R>20.0$. Further observations reported the nova to be at
unfiltered magnitudes of 18.4 on Oct 22.99, 19.1 on 23.43, and $>19.7$ on Oct 24.47. K. Hornoch
reported an $R$-band magnitude of $18.18\pm0.08$
on Oct
23.12\footnote{http://www.cbat.eps.harvard.edu/unconf/followups/J00452885+4154094.html}.
On Oct 26.97, the nova had a $B$-band magnitude of $20.9\pm0.15$ and 
$V=21.1\pm0.16$ \citep{2011ATel.3725....1B}.

\begin{figure}
\includegraphics[width=\columnwidth]{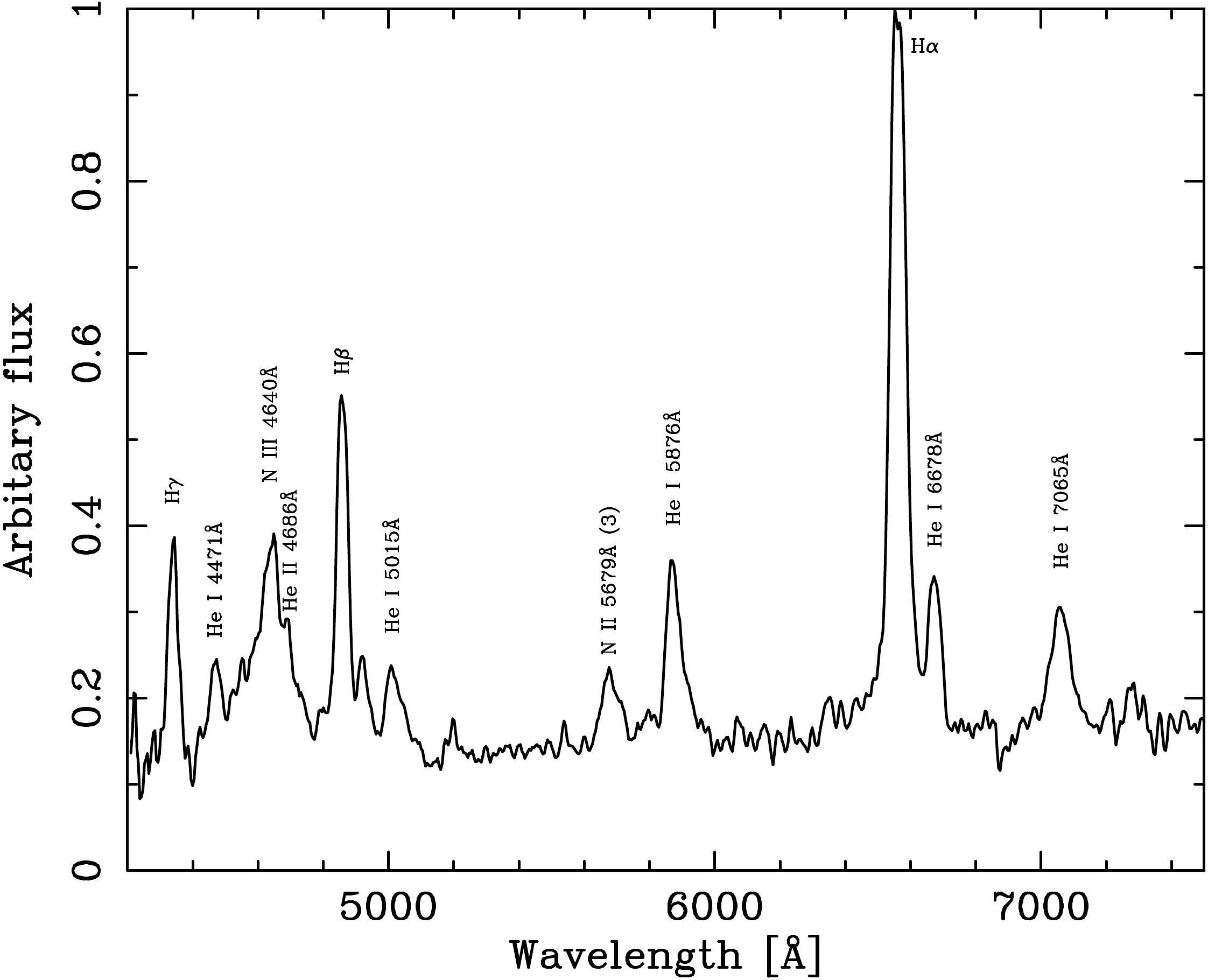}
\caption{\label{spec}Hobby-Eberly Telescope LRS spectrum of
  \object{M31N 2008-12a} (truncated at $7,500$\AA)
  taken on 2012 Oct 20.34 UT, $\sim2$ days after the 2012 eruption
  (\object{M31N 2012-10a}).}
\end{figure}

In 2012, another eruption
(\object{M31N 2012-10a}) was
discovered at $0^{\mathrm{h}}45^{\mathrm{m}}28^{\mathrm{s}}\!.84$,
$+41^{\circ}54^{\prime}09^{\prime\prime}\!\!.5$ by K. Nishiyama and
F. Kabashima. They measured the nova to be at an unfiltered magnitude of
18.9 on 2012 Oct 18.68 UT. The object, which was not visible to
a limiting magnitude of 19.8 on Oct 15.52, appeared to brighten to
18.6 by Oct 19.51\footnote{http://www.cbat.eps.harvard.edu/unconf/followups/J00452884+4154095.html}.
The nova was observed at $R=18.45\pm0.04$ on
Oct 19.72 and $i'=18.42\pm0.06$ on Oct 19.73
\citep{2012ATel.4503....1S}. A spectrum of the transient was taken by
\citet[see also Fig.~\ref{spec}]{2012ATel.4503....1S} on Oct 20.34, which was consistent with
that of a He/N nova in \object{M31}.  The FWHM of the Balmer emission lines
indicated an ejecta expansion velocity of $2,250\;\mathrm{km/s}$.  

\section{Observations of the 2013 Eruption}

\begin{figure}
\includegraphics[width=\columnwidth]{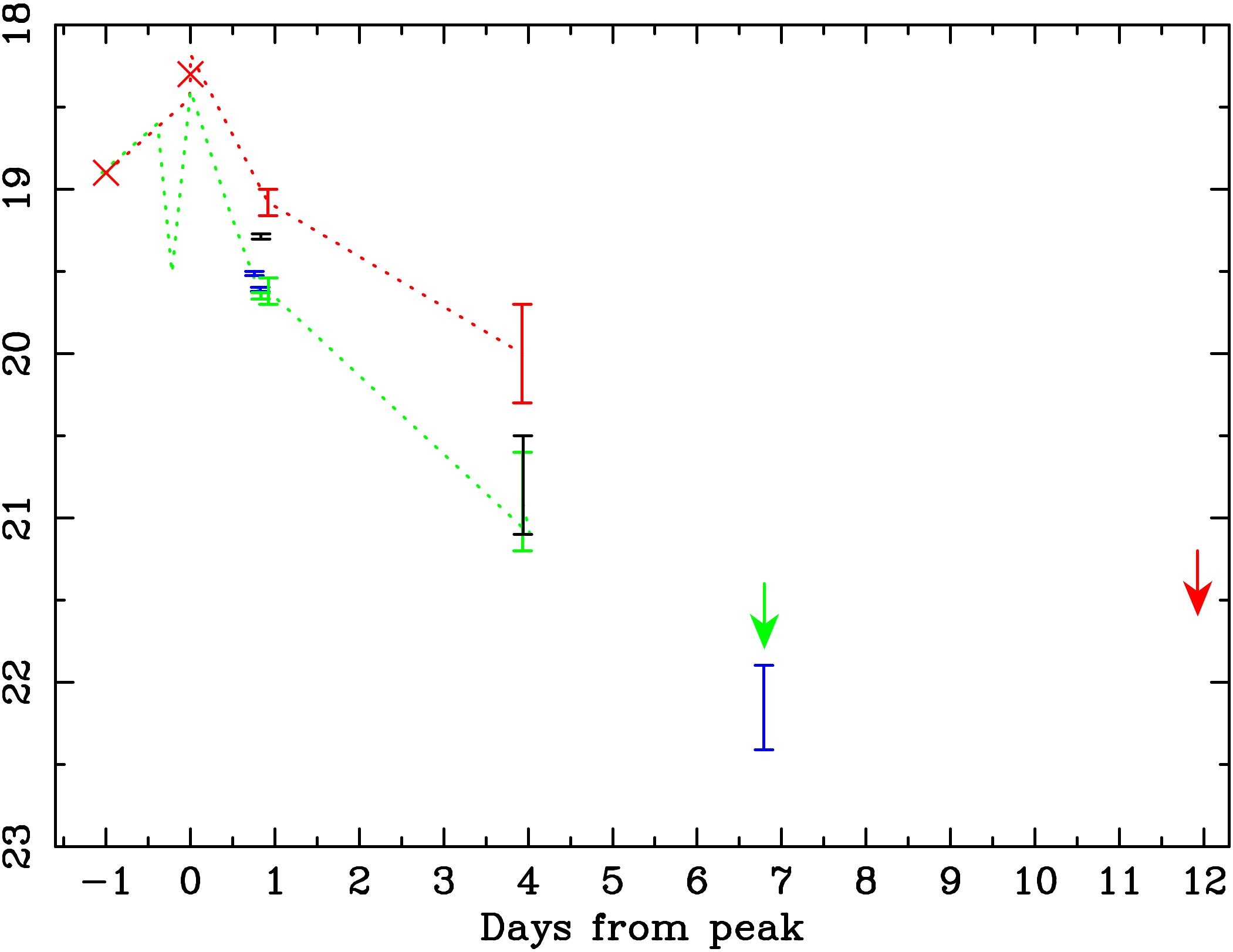}
\caption{\label{2013lightcurve}Optical lightcurve of the 2013 eruption
  of \object{M31N 2008-12a}, data from \protect{\citet[$\times$
    symbols]{ATel5607}}, LT (3rd and 5th epochs), and Danish 1.54m
  (4th and 6th epochs).  Blue data: $B$,
  green: $V$, red: $R$ and black: $I$/$i'$.  Dotted lines indicate a
  template `generic' lightcurve based on the $V$ (green) and $R$ (red)
observations of the 2008, 11, 12 and 13 eruptions.}
\end{figure}

The 2013 eruption (\object{M31N 2013-11f}) was discovered by the intermediate Palomar Transient
Factor (iPTF) on 2013 Nov 27.1 UT (ID: \object{PTF09hsd})
at $0^{\mathrm{h}}45^{\mathrm{m}}28^{\mathrm{s}}\!.89$,
$+41^{\circ}54^{\prime}10^{\prime\prime}\!\!.2$, subsequently
peaking at $R=18.3$ on Nov 28.1 \citep{ATel5607}.  Broadband $B$, $V$ and $i'$ photometry
was obtained with the IO:O CCD camera on the LT approximately one and
seven days after peak.  LT observations are part
of a larger programme of photometry and spectroscopy of novae in \object{M31}
\citep[see for example,][]{2011ApJ...734...12S}.  Photometric observations were also obtained using the Danish 1.54m telescope
at La Silla four and six days after maximum.  These data were reduced using standard
routines within IRAF \citep{1993ASPC...52..173T}, and
calibrated against secondary standards in \object{M31}
\citep{2006AJ....131.2478M}.
The photometry is reported in
Table~\ref{obs_tab} and the 2013 lightcurve is presented in
Fig.~\ref{2013lightcurve}.

\begin{table}
\caption{Observations of the 2013 eruption of \object{M31N 2008-12a} and
  archival {\it HST} observations covering the position of the progenitor.}
\label{obs_tab}
\begin{center}
\begin{tabular}{lll}
\hline\hline
JD & Telescope \& & Photometry \\
2450000+ & Instrument & \& Filter \\
\hline
5416.027 & {\it HST} ACS/WFC\tablefootmark{a,b} & $\mathrm{F475W}=24.07\pm0.02$ \\
5415.956 & {\it HST} ACS/WFC\tablefootmark{a,b} & $\mathrm{F814W}=23.90\pm0.02$ \\
\hline
5586.715 & {\it HST} WFC3/UVIS\tablefootmark{b} & $\mathrm{F275W}=23.14\pm0.06$ \\
5586.705 & {\it HST} WFC3/UVIS\tablefootmark{b} & $\mathrm{F336W}=23.10\pm0.03$ \\
5586.771 & {\it HST} WFC3/IR\tablefootmark{b} & $\mathrm{F110W}>22.05$ \\
5586.780 & {\it HST} WFC3/IR\tablefootmark{b} & $\mathrm{F160W}>21.22$ \\
\hline
5805.013 & {\it HST} WFC3/UVIS\tablefootmark{c} & $\mathrm{F275W}=22.9\pm0.1$ \\
5805.012 & {\it HST} WFC3/UVIS\tablefootmark{c} & $\mathrm{F336W}=22.81\pm0.03$ \\
5805.079 & {\it HST} WFC3/IR\tablefootmark{c} & $\mathrm{F110W}>21.01$ \\
5805.087 & {\it HST} WFC3/IR\tablefootmark{c} & $\mathrm{F160W}>21.18$ \\
\hline
5936.631 & {\it HST} ACS/WFC\tablefootmark{c} & $\mathrm{F475W}=24.49\pm0.02$ \\ 
5936.538 & {\it HST} ACS/WFC\tablefootmark{c} & $\mathrm{F814W}=24.05\pm0.02$ \\
\hline
6625.357 & LT IO:O\tablefootmark{a} & $B=19.51\pm0.01$ \\
6625.425 & LT IO:O\tablefootmark{a} & $B=19.61\pm0.01$ \\
6625.430 & LT IO:O\tablefootmark{a} & $V=19.65\pm0.02$ \\
6625.435 & LT IO:O\tablefootmark{a} & $i'=19.29\pm0.02$ \\
6625.519 & Danish 1.54m DFOSC & $R=19.08\pm0.08$ \\
6625.522 & Danish 1.54m DFOSC & $V=19.62\pm0.09$ \\
6628.529 & Danish 1.54m DFOSC & $R=20.0\pm0.3$ \\
6628.533 & Danish 1.54m DFOSC & $V=20.9\pm0.3$ \\
6628.537 & Danish 1.54m DFOSC & $I=20.8\pm0.3$ \\
6631.391 & LT IO:O & $B=22.2\pm0.3$ \\
6631.396 & LT IO:O & $V>21.4$ \\
6636.528 & Danish 1.54m DFOSC & $R>21.2$ \\
\hline
\end{tabular}
\end{center}
\tablefoot{
\tablefoottext{a}{\citet{ATel5611}}
\tablefoottext{b}{Prop. ID: 12056}
\tablefoottext{c}{Prop. ID: 12106}
}
\end{table}

With limited optical coverage of each eruption, we make the assumption that all eruptions of a RN are essentially
similar \citep{2010ApJS..187..275S}.  We have constructed single $R$- and $V$-band lightcurves of the
eruption using data from the 2008, 11, 12 and 13 eruptions (see
Table~\ref{obs_tab} and Introduction text).  These `generic'
lightcurves are represented by the {\it dotted} lines in
Fig.~\ref{2013lightcurve}.  Based on these generic lightcurves, we
estimate that the decline times of this RN are
$t_{2}\left(V\right)\simeq4\;\mathrm{days}$ and
$t_{2}\left(R\right)\simeq5\;\mathrm{days}$ classifying this RN as {\it
  very fast}.  The peak magnitudes observed over these four eruptions
are $V=18.4$ and $R=18.18$.

The astrometric position of the 2013 eruption was measured from an LT
$i'$-band image taken on 2013 Nov 28.94 UT.  An astrometric solution
was obtained using 14 stars from the Two Micron All Sky Survey (2MASS)
All-Sky Catalogue \citep{2006AJ....131.1163S} which are coincident with resolved
sources in the LT observation.  We obtain a position for the 2013
eruption of
$\alpha=0^{\mathrm{h}}45^{\mathrm{m}}28^{\mathrm{s}}\!.82\pm0^{\mathrm{s}}\!.01$,
$\delta=+41^{\circ}54^{\prime}10^{\prime\prime}\!\!.1\pm0^{\prime\prime}\!\!.1$
(the astrometric uncertainty
is dominated by uncertainties in the plate solution).

A target of opportunity monitoring campaign was also initiated with
the {\it Swift} satellite, see \citet{ATel5627,ATel5633} and HND2014 for a full
discussion.

\section{Progenitor System}

Following the procedure outlined in
\citet{2009ApJ...705.1056B} and Williams et~al. (2014, sub.), we undertook a search for any resolved progenitor
system of the 2013 eruption within positionally coincident archival {\it Hubble Space
  Telescope (HST)} data{\footnote{Data taken by the Panchromatic
  Hubble Andromeda Treasury survey \citep[see, for example,][]{2012ApJS..200...18D}}.  The position of the 2013 eruption of the RN
\object{M31N 2008-12a} was isolated within the archival {\it HST} data by
calculating a geometric spatial transformation between the LT
($i'$-band; one day post maximum) and
{\it HST} ACS/WFC F814W data, a method that is independent of the absolute
astrometric calibration of the data.  This was performed using 16
stars that were visible and unsaturated in both datasets. 

There is a resolved object 0.556 ACS/WFC pixels from the position of
the 2013 eruption in the {\it HST} ACS/WFC F814W  image taken on 2010 Aug 7
(from proposal ID: 12056; see Fig.~\ref{LTimage}).  This represents a separation of 28
milli-arcseconds ($0.9\sigma$) from the eruption position.  The
probability of finding an object at least as close to the eruption
position, based on the local resolved stellar density, is only
$2.5\%$.  Hence we can be very confident that the object in the {\it
  HST} data is related to the nova eruption and likely the progenitor/quiescent system. 

\begin{figure}
\includegraphics[width=\columnwidth]{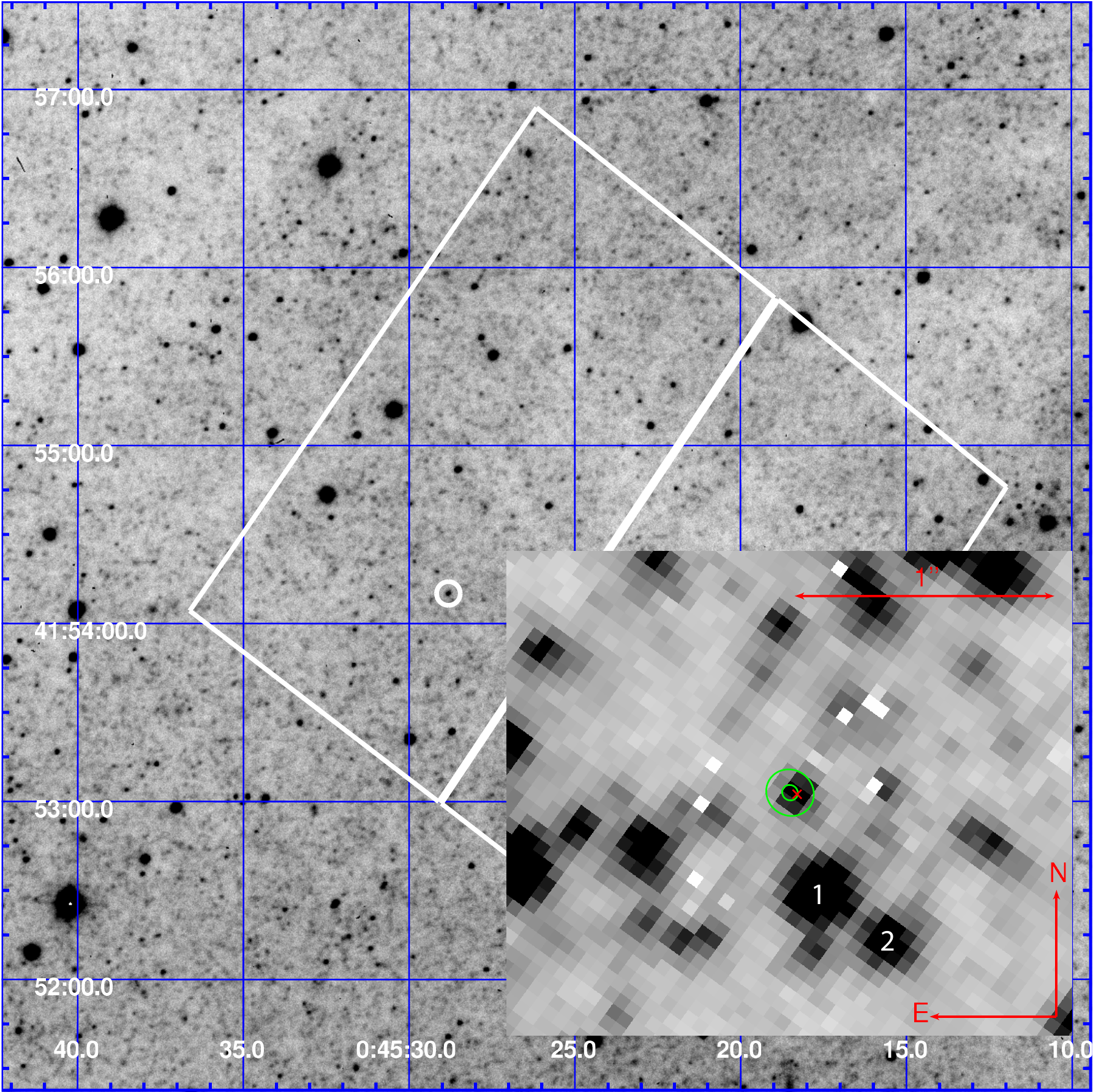}
\caption{\label{LTimage}LT $i'$-band image of
  the 2013 eruption of \object{M31N 2008-12a} taken on 2013 Nov 28.94
  UT.  The RN position is
  shown by the white circle, the white boxes indicate the
  coincident {\it HST} ACS/WFC fields.  Inset:
  {\it HST} ACS/WFC F814W image of the
  $\sim2^{\prime\prime}\times2^{\prime\prime}$ region surrounding
  \object{M31N 2008-12a}.  The inner and outer green ellipses indicate the $1\sigma$
  (31 mas) and $3\sigma$ radius progenitor search regions respectively, and the red cross indicates
  the position of the progenitor candidate.  See text for discussion
  regarding stars 1 and 2.}
\end{figure}

{\it HST} archival data were available from two proposals (12056 and
12106), both provided optical F475W and F814W data using ACS/WFC, near-UV
F275W and F336W data using WFC3/UVIS, and near-IR F110W and F160W data
using WFC3/IR. Photometry of these data was
undertaken using DOLPHOT \citep[v2.0;][following the standard procedure and parameters given in the
manual]{2000PASP..112.1383D}.  The photometry of the candidate
progenitor system is reported in
Table~\ref{obs_tab}.  Whilst the candidate progenitor system was
resolved in the optical and NUV {\it HST} data, any object at the
eruption position would have been severely blended with star 1 (see
Fig.~\ref{LTimage}) in the NIR data.  Therefore we present, as
upper limits on the F110W and F160W magnitudes of the candidate
progenitor systen, the NIR photometry of star 2.  As star 2 is
just resolvable from star 1 and is marginally closer to this bright
star than the RN these represent conservative upper limits on the
progenitor system brightness.

With multiple waveband observations available a
meaningful distance and extinction-corrected spectral energy distribution (SED) can be
produced for the quiescent \object{M31N 2008-12a} which can be directly
compared to that of known Galactic RNe.  Fig.~\ref{SED} presents the
SED of the quiescent \object{M31N 2008-12a} and those of the RG-novae
\object{RS Oph}
and \object{T CrB} and the SG-nova \object{U Sco}, we have followed the methodology
outlined in \citet{2010ApJS..187..275S} to allow direct comparison
with their Galactic RN SED (see their Figure~71).  Quiescent photometry for the
Galactic RNe is taken from \citet[see their
Table~30]{2010ApJS..187..275S}, distances and extinction from
\citet[see their Table~2 and references therein]{2012ApJ...746...61D},
optical and NIR absolute calibrations from
\citet{1979PASP...91..589B} and \citet{1985AJ.....90..896C}
respectively.  It should be noted that here we use a significantly
different (closer) distance to \object{RS Oph}, $1.4^{+0.6}_{-0.2}$~kpc
(\citealp{2008ASPC..401...52B}; see also
\citealp{1987rorn.conf..241B}).  For \object{M31N 2008-12a} we
assume a distance to \object{M31} of $770\pm19$~kpc
\citep{1990ApJ...365..186F}, a line-of-sight external (Galactic)
reddening of $E_{B-V}=0.1$ \citep{1992ApJS...79...77S} and additional
internal (\object{M31}) reddening of $E_{B-V}\leq0.16$ \citep{2009A&A...507..283M}.

\begin{figure}
\includegraphics[width=\columnwidth]{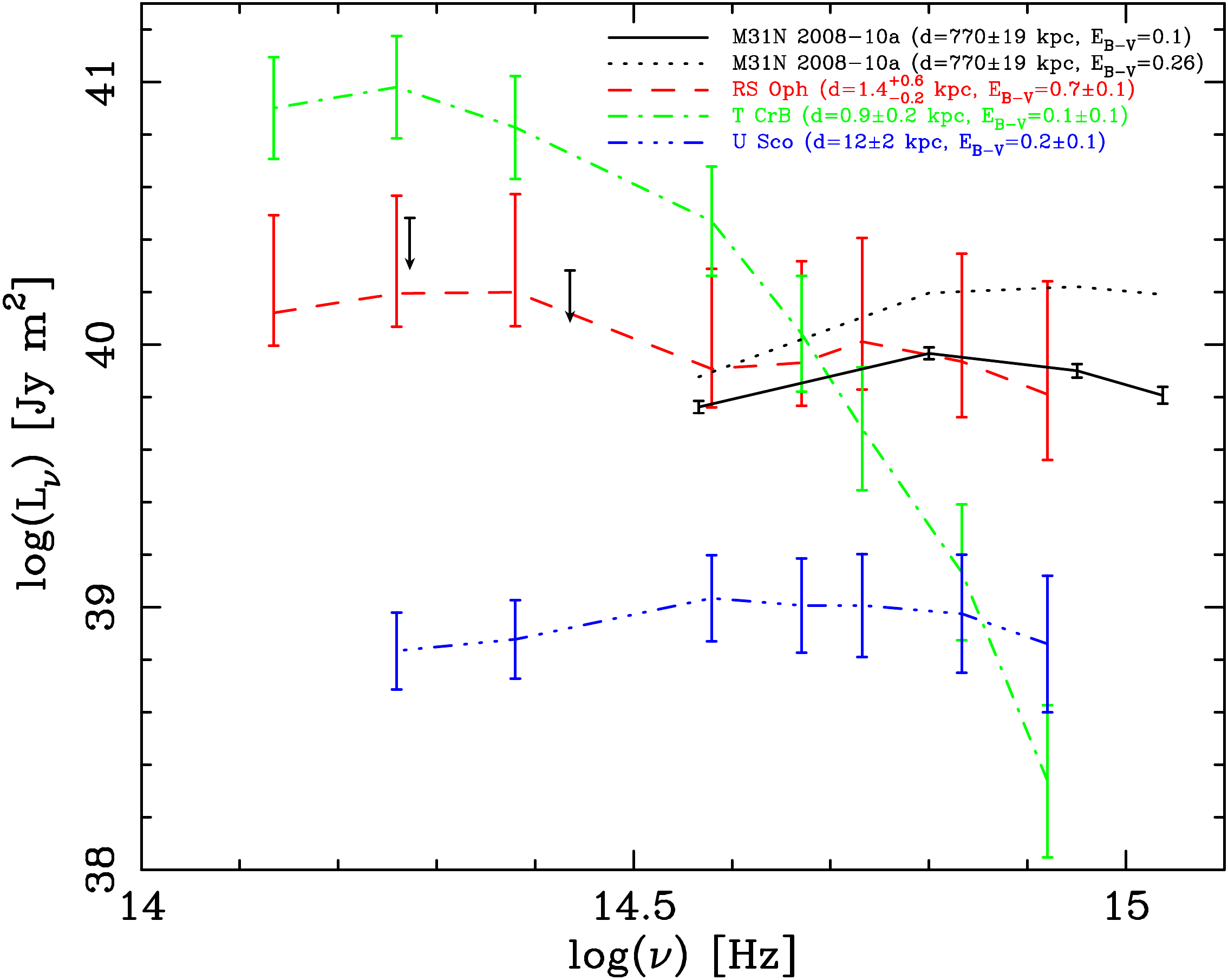}
\caption{\label{SED}Distance and extinction-corrected SEDs for the
  progenitor of \object{M31N 2008-12a} compared to those
  of the quiescent RNe \object{RS Oph}, \object{T CrB} 
  and \object{U Sco}  (see Key for line identifications).  Units chosen to allow comparison with a
similar plot in \protect{\citet[see their
  Figure~71]{2010ApJS..187..275S}}.  The black dotted line indicates
the maximum effect of any internal \object{M31} extinction ($E_{B-V}^{\mathrm{internal}}\leq0.16$).  For each
system, point-to-point uncertainties are small, indicated error bars are
dominated by distance and extinction uncertainties.}
\end{figure}

\section{Discussion}

With three eruptions over a two year period (2011, 12 and 13), the RN
\object{M31N 2008-12a} is a
unique system.  Such a short ($\sim1$~year) recurrence time can be
expected from a system with a low critical mass for ignition, which requires
a high-mass WD.  Further, to accumulate enough mass for ignition within
a short time, a high mass accretion rate is needed.  Nova evolution models
published by \citet{2005ApJ...623..398Y} indicate that such extremely
short recurrence times are in fact possible but require both a
high-mass WD, close to $1.4\mathrm{M}_{\odot}$, and a high mass accretion
rate ($-8<\log \mathrm{\dot{M}}/\mathrm{M}_{\odot}<-7$, see their
Table~3 and discussion in HND2014).  However, the relatively low
optical luminosity
($V_{\mathrm{max}}=18.4$) and moderate `ejecta' velocity
($2,250\;\mathrm{km/s}$) are slightly puzzling.  The former is
discussed in more detail in HND2014 and the derived velocity may be in
part due to the inclination of highly shaped ejecta (as may be
expected in the presence of a massive accretion disk).

Given the eruptions in 2011, 12 and 13, each separated by approximately
a year, and eruptions in  2008 and 09 \citep{2014arXiv1401.2426T} it
seems likely that \object{M31N 2008-12a} has a recurrence timescale of
$\sim1$ year and that an eruption towards the end of 2010 was missed.  HND2014 report on the
subsequent X-ray detection following the 2013 eruption but they also
summarise previous X-ray detections at a similar position.  Transient
X-ray sources were also detected in early 1992 and 93 and in Sep
2001, indicating that this system may have been experiencing yearly
eruptions for at least 20 years.  The relative faintness of this
eruption, its very rapid decline and its position far out in the disk
of \object{M31} may all account for a high number of `missed' eruptions.

With so many eruptions in such a short time, we must address whether
these could be due to spatial coincidence.  Following the procedure in \citet[see their
Eq.~6]{2011ApJ...734...12S}, the probability of a chance positional
coincidence at the location of the nova and within the error circle
defined by the reported positions of the 2008, 11 and 12 eruptions is just 0.0002.

The distance and extinction-corrected SED of the quiescent
\object{M31N 2008-12a} (see Fig.~\ref{SED}) is remarkably similar in the
optical to the Galactic RN \object{RS Oph}.  The SED of \object{RS Oph}, with its short
($\sim20$ year)
inter-eruption period, is a combination
of the RGB secondary (NIR) and the accretion disk (optical
and NUV), unlike that of \object{T CrB}, with a longer recurrence time
($\sim80$ years),
where the SED is dominated by the RGB secondary and there is
little sign of a disk.  The SEDs of SG-novae, e.g.\ \object{U Sco}, are dominated by the accretion disk with little or no
contribution from the less evolved, less luminous, secondary.  Given
the form and luminosity of the \object{M31N 2008-12a} progenitor SED it is likely that the progenitor of
\object{M31N 2008-12a} also contains a significant accretion disk that
dominates the NUV and optical flux.  

While, based on the SED, a RG-nova system similar to
\object{RS Oph} seems
the most likely scenario, a SG-nova system (akin to \object{U Sco}) may still
be possible.  The SED of \object{U Sco} is dominated by its accretion disk, but
as \object{U Sco} is an eclipsing system the disk is observed edge-on, i.e.\ at its
faintest.  Given the short recurrence time of \object{M31N 2008-12a}, the
observed SED could be due solely to an extremely bright (i.e.\ very high
accretion rate) almost face-on accretion disk.  In order to confirm
the evolutionary nature of the secondary, stronger limits (or a
detection) are needed in the NIR bands, requiring deeper or higher
spatial resolution images.  Alternatively, the secondary nature could
be inferred if the orbital period or inclination of the system can be determined.

The {\it HST} archival data is separated into four epochs; the ACS/WFC
observations from proposal IDs 12056 and 12106 were taken in Aug
2010 ($\sim14$ months before the 2011 eruption) and Jan 2012
($\sim3$ months after the 2011 eruption and $\sim9$ months before the
2012 eruption), respectively.  The WFC3 observations were taken in
Jan 2011 and Aug 2011, $\sim9$ and $\sim2$ months before the
2011 eruption, respectively.  Given the rapid ($t_{2}\left(V\right)\simeq4$ days) decline time of the
eruption of \object{M31N 2008-12a}, all the {\it HST} observations are
sufficiently distant from any reported eruptions that the system is
likely to be near or at quiescence during these observations.  That
is, we are unlikely to be observing the late decline of any eruption in
the {\it HST} data.  The similarity between the photometry from the
two {\it HST} datasets implies that even if an eruption in
Sep-Dec 2010 has been missed the system was back at
quiescence by the end of Jan 2011.

\section{Conclusions}

The RN \object{M31N 2008-12a} has had five recorded eruptions in the past five
years, in 2008, 09, 11, 12 and 13.  Combined data from four of these
eruptions indicate a very fast He/N nova with a decline time $t_{2}\left(V\right)\simeq4$
days.  These observations, coupled with transient X-ray
detections in 1992, 93 and 2001, indicate that this system has a
remarkably short $\sim1$ year recurrence time.  This points to a
system containing a very high mass WD with a high accretion rate.  A
search of archival {\it HST} data indicates a candidate progenitor
system, most likely containing a RGB secondary (RG-nova) and bright
accretion disk (e.g.\ RS~Oph), although a SG-nova progenitor
(e.g.\ U~Sco) can't be ruled out.

In addition to this Letter, HND2014 report on the X-ray observations
of \object{M31N 2008-12a} and a follow-up paper will study the optical and
X-ray archives in more detail.  \object{M31N 2008-12a} is a unique system and
we encourage further observations, particularly towards the end of 2014.

\begin{acknowledgements}
The LT
is operated on the island of La Palma by LJMU in the Spanish Observatorio del Roque de
los Muchachos of the Instituto de Astrofisica de Canarias with
financial support from STFC.  Based (in part) on data collected with the Danish 1.54-m telescope at the ESO
La Silla Observatory.  SCW acknowledges PhD funding from STFC.
MH acknowledges support from an ESA fellowship. AWS acknowledges
support from NSF grant AST1009566.  Work of KH and VV was supported by the project
RVO:67985815 and by the grant LG12001 of the Ministry of Education of the Czech
Republic.  Finally, we thank an anonymous referee for their
constructive suggestions.
\end{acknowledgements}


\begin{thebibliography}{}

\bibitem[Barry et al.(2008)]{2008ASPC..401...52B} Barry, R.~K., Mukai, K., 
Sokoloski, J.~L., et al.\ 2008, in ASP Conf. Ser. 401, RS Ophiuchi
(2006) and the Recurrent Nova Phenomenon, ed. A. Evans, M.~F. Bode,
T.~J. O’Brien, \& M.~J. Darnley (San Francisco, CA: ASP), 52 

\bibitem[Barsukova et al.(2011)]{2011ATel.3725....1B} Barsukova, E., 
Fabrika, S., Hornoch, K., et al.\ 2011, ATel,
3725 

\bibitem[Bessell(1979)]{1979PASP...91..589B} Bessell, M.~S.\ 1979, \pasp, 
91, 589 

\bibitem[Bode(1987)]{1987rorn.conf..241B} Bode, M.~F.\ 1987, RS Ophiuchi 
(1985) and the Recurrent Nova Phenomenon, ed. M.~F. Bode (Utrecht: VNU
Science), 241 

\bibitem[Bode 
\& Evans(2008)]{2008clno.book.....B} Bode, M.~F., \& Evans, A.\ 2008,
Classical Novae, 2nd Edition.~Edited by M.F.~Bode and
A.~Evans.~Cambridge Astrophysics Series, No.~43, Cambridge
University Press  

\bibitem[Bode et al.(2009)]{2009ApJ...705.1056B} Bode, M.~F., Darnley, 
M.~J., Shafter, A.~W., et al.\ 2009, \apj, 705, 1056 

\bibitem[Campins et al.(1985)]{1985AJ.....90..896C} Campins, H., Rieke, 
G.~H., \& Lebofsky, M.~J.\ 1985, \aj, 90, 896 

\bibitem[Dalcanton et al.(2012)]{2012ApJS..200...18D} Dalcanton, J.~J., 
Williams, B.~F., Lang, D., et al.\ 2012, \apjs, 200, 18 

\bibitem[Darnley et al.(2006)]{2006MNRAS.369..257D} Darnley, M.~J., Bode, 
M.~F., Kerins, E., et al.\ 2006, \mnras, 369, 257 

\bibitem[Darnley et al.(2012)]{2012ApJ...746...61D} Darnley, M.~J., 
Ribeiro, V.~A.~R.~M., Bode, M.~F., Hounsell, R.~A., 
\& Williams, R.~P.\ 2012, \apj, 746, 61 

\bibitem[Dolphin(2000)]{2000PASP..112.1383D} Dolphin, A.~E.\ 2000, \pasp, 
112, 1383 

\bibitem[Freedman 
\& Madore(1990)]{1990ApJ...365..186F} Freedman, W.~L., \& Madore,
B.~F.\ 1990, \apj, 365, 186 

\bibitem[Henze et al.(2013a)]{ATel5627} Henze, M., Ness, J.-U., Bode,
  M.~F., Darnley, M.~J., Williams, S.~C.\ 2013a, ATel, 5627 

\bibitem[Henze et al.(2013b)]{ATel5633} Henze, M., Ness, J.-U., Bode,
  M.~F., et al.\ 2013b, ATel, 5633 

\bibitem[Henze et al.(2014)]{2014arXiv1401.2904H} Henze, M., Ness, J.-U., 
Darnley, M.~J., et al.\ 2014, \aap, in press, arXiv:1401.2904
(HND2014) 

\bibitem[Massey et al.(2006)]{2006AJ....131.2478M} Massey, P., Olsen, 
K.~A.~G., Hodge, P.~W., et al.\ 2006, \aj, 131, 2478 

\bibitem[Montalto et 
al.(2009)]{2009A&A...507..283M} Montalto, M., Seitz, S., Riffeser, A.,
et al.\ 2009, \aap, 507, 283 

\bibitem[Pietsch et 
al.(2007)]{2007A&A...465..375P} Pietsch, W., Haberl, F., Sala, G., et
al.\ 2007, \aap, 465, 375 

\bibitem[Schaefer(2010)]{2010ApJS..187..275S} Schaefer, B.~E.\ 2010, \apjs, 
187, 275  

\bibitem[Shafter et al.(2011)]{2011ApJ...734...12S} Shafter, A.~W., 
Darnley, M.~J., Hornoch, K., et al.\ 2011, \apj, 734, 12 

\bibitem[Shafter et al.(2012)]{2012ATel.4503....1S} Shafter, A.~W., 
Hornoch, K., Ciardullo, J.~V.~R., Darnley, M.~J., 
\& Bode, M.~F.\ 2012, ATel, 4503 

\bibitem[Skrutskie et al.(2006)]{2006AJ....131.1163S} Skrutskie, M.~F., 
Cutri, R.~M., Stiening, R., et al.\ 2006, \aj, 131, 1163 

\bibitem[Stark et al.(1992)]{1992ApJS...79...77S} Stark, A.~A., Gammie, 
C.~F., Wilson, R.~W., et al.\ 1992, \apjs, 79, 77 


\bibitem[Steele et al.(2004)]{2004SPIE.5489..679S} Steele, I.~A., Smith, 
R.~J., Rees, P.~C., et al.\ 2004, \procspie, 5489, 679 

\bibitem[Tang et al.(2013)]{ATel5607} Tang, S., Cao, Y., Kasliwal,
  M.~M.\ 2013, ATel, 5607 

\bibitem[Tang et al.(2014)]{2014arXiv1401.2426T} Tang, S., Bildsten, L., 
Wolf, W.~M., et al.\ 2014, \apj, submitted, arXiv:1401.2426 

\bibitem[Tody(1993)]{1993ASPC...52..173T} Tody, D.\ 1993, Astronomical Data 
Analysis Software and Systems II, 52, 173 

\bibitem[Williams et al.(2013)]{ATel5611} Williams, S.~C., Darnley,
  M.~J., Bode, M.~F., Shafter, A.~W.\ 2013, ATel,
  5611 

\bibitem[Yaron et al.(2005)]{2005ApJ...623..398Y} Yaron, O., Prialnik, D., 
Shara, M.~M., \& Kovetz, A.\ 2005, \apj, 623, 398 

\end{thebibliography}
\end{document}